\documentclass{article}
\usepackage{LaThuileFPSpro}
\begin{document}
\title{HIGH PRECISION STUDY OF CP-VIOLATING CHARGE ASYMMETRY
       IN $K^\pm\to3\pi^\pm$ DECAYS}
\author{
  Evgueni Goudzovski\\
  {\em Joint Institute for Nuclear Research, Dubna, Russia} \\
  on behalf of the NA48/2 collaboration
  }
\maketitle

\baselineskip=11.6pt

\begin{abstract}
A precise measurement of the direct CP violating charge asymmetry
par\-am\-et\-er $A_g$ in $K^\pm\to\pi^\pm\pi^+\pi^-$ decays by the
NA48/2 experiment at CERN SPS is presented. The experiment has been
designed not to be limited by systematic uncertainties in the
asymmetry measurement. A preliminary result for the charge asymmetry
$A_g=(-1.3\pm2.3)\times 10^{-4}$ has been obtained with a sample of
$3.11\times 10^9$ selected events corresponding to the full
collected statistics. The precision of the result is limited by the
statistics used.
\end{abstract}
\newpage
\section*{Introduction}
More than 40 years after its discovery\cite{ch64}, the phenomenon of
CP violation still plays a central role in present and future
particle physics investigations. For a long time the effect seemed
to be confined to a peculiar sector of particle physics. However,
two important discoveries took place recently. In the late 1990's,
following an earlier indication by NA31\cite{ba93}, the NA48 and
KTeV experiments firmly established the existence of direct CP
violation\cite{fa99,al99} by measuring a non-zero
$\varepsilon'/\varepsilon$ parameter in the decays of neutral kaons
into two pions. During the last five years, the $B$-factory
experiments Babar and Belle have discovered a series of
indirect\cite{au01} and direct\cite{ab04} CP violating effects in
the system of neutral $B$ mesons.

In order to explore possible non-Standard Model (SM) enhancements to
heavy-quark loops, which are at the core of direct CP-violating
processes, all manifestations of direct CP violation have to be
studied experimentally. In kaon physics, besides the parameter
$\varepsilon'/\varepsilon$ already measured in $K_{L,S}\to2\pi$
decays, the most promising complementary observables are decay rates
of GIM-suppressed flavour-changing neutral current decays
$K\to\pi\nu\bar\nu$ and the charge asymmetry between $K^+$ and $K^-$
decays into three pions.

Direct CP violation in decay amplitude (as in the $K^\pm\to3\pi$
case) provides a strong qualitative test of the way the SM
accommodates CP violation. However, its quantitative exploration to
constrain the fundamental parameters of the theory is difficult due
to non-perturbative hadronic effects. Still, an intense theoretical
programme is under way to improve such predictions, which will
ultimately allow the direct CP violation measurements to be used as
strong quantitative constraints on the SM.

The $K^\pm\to3\pi$ matrix element squared is conventionally
parameterized by a polynomial expansion\cite{pdg}
\begin{equation}
|M(u,v)|^2\sim 1+gu+hu^2+kv^2, \label{slopes}
\end{equation}
where $g$, $h$, $k$ are the so called linear and quadratic Dalitz
plot slope parameters ($|h|,|k|\ll |g|$) and the two Lorentz
invariant kinematic variables $u$ and $v$ are defined as
\begin{equation}
u=\frac{s_3-s_0}{m_\pi^2},~~v=\frac{s_2-s_1}{m_\pi^2},~~
s_i=(P_K-P_i)^2,~i=1,2,3;~~s_0=\frac{s_1+s_2+s_3}{3}.
\end{equation}
Here $m_\pi$ is the charged pion mass, $P_K$ and $P_i$ are the kaon
and pion four-momenta, the indices $i=1,2$ correspond to the two
identical (``even'') pions and the index $i=3$ to the pion of
different charge (the ``odd'' pion). A term linear in $v$ is
forbidden in (\ref{slopes}) due to symmetry considerations. A
difference of slope parameters $g^+$ and $g^-$ describing positive
and negative kaon decays, respectively, is a manifestation of direct
CP violation usually defined by the corresponding slope asymmetry
\begin{equation}
A_g = (g^+ - g^-)/(g^+ + g^-) \approx \Delta g/(2g),
\end{equation}
where $\Delta g$ is the slope difference and $g$ is the average
slope. The asymmetry of integrated decay rates is expected to be
strongly suppressed with respect to the slope asymmetry\cite{is92}.
The SM predictions for $A_g$\cite{ma95} vary from a few $10^{-6}$ to
a few $10^{-5}$. Existing theoretical calculations involving
processes beyond the SM\cite{sh98} allow a wider range of $A_g$,
including substantial enhancements up to a few $10^{-4}$. Several
experiments have searched for the asymmetry $A_g$ in both
$\pi^\pm\pi^+\pi^-$ and $\pi^\pm\pi^0\pi^0$ decay modes\cite{fo70}.
The upper limits reached so far are at the level of a few $10^{-3}$,
and are limited by systematic uncertainties.

The NA48/2 experiment is carried out in the framework of kaon
physics programme at the CERN SPS. Its primary aim is to measure the
asymmetries $A_g$ and $A_g^0$ in $K^\pm\to\pi^\pm\pi^+\pi^-$ and
$K^\pm\to\pi^\pm\pi^0\pi^0$ decays, respectively, with a precision
at least one order of magnitude better than the existing limits,
which would significantly reduce the existing gap between
experiments and theory.

A measurement of $A_g$ performed with approximately a half of the
NA48/2 data sample has been published\cite{k3pi}. The current paper
presents a preliminary result based on the full data sample.
\section{Beams and detectors}
\begin{figure}[t]
  \includegraphics{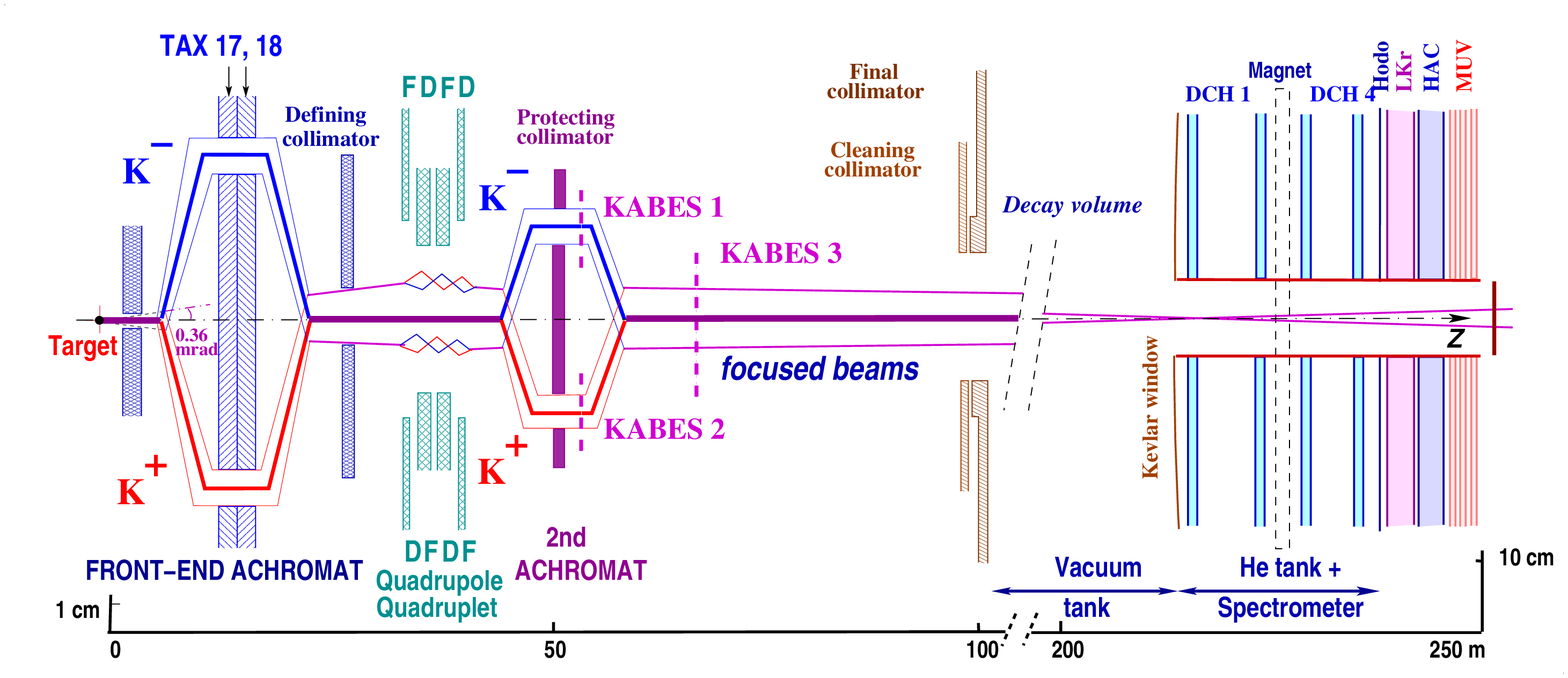}
    \vspace{50mm}
  \caption{\it
    Schematic lateral view of the NA48/2 beam
line (TAX17,18: motorized beam dump/collimators used to select the
momentum of the $K^+$ and $K^-$ beams; DFDF: focusing quadrupoles,
KABES1--3: kaon beam spectrometer stations), decay volume and
detector (DCH1--4: drift chambers, Hodo: hodoscope, LKr: EM
calorimeter, HAC: hadron calorimeter, MUV: muon veto). The vertical
scale changes for the two parts of the figure.
    \label{fig:beams} }
\end{figure}

High precision measurement of $A_g$ (at the level of $10^{-4}$)
requires not only high statistics, but also a dedicated experimental
approach. A beam line transporting two simultaneous charged beams of
opposite signs was designed and built as a key element leading to
cancellations of main systematic uncertainties, allowing decays of
$K^+$ and $K^-$ to be recorded at the same time. Regular alternation
of magnetic fields in all the beam line elements and the
spectrometer magnet was adopted to symmetrize the acceptance for the
two beams. The layout of the beams and detectors is shown
schematically in Fig.~\ref{fig:beams}.

The setup is described in a right-handed orthogonal coordinate
system with the $z$ axis directed downstream along the beam, and the
$y$ axis directed vertically up.

The beams are produced by 400 GeV protons impinging on a beryllium
target of 40 cm length and 2 mm diameter at zero incidence angle.
Charged particles with momentum $(60\pm3)$ GeV/$c$ are selected
symmetrically for positive and negative particles by an achromatic
system of four dipole magnets with null total deflection, which
splits the two beams in the vertical plane and then recombines them
on a common axis. They then pass through a defining collimator and a
series of four quadrupoles designed to produce horizontal and
vertical charge-symmetric focusing of the beams towards the
detector. Finally they are again split and recombined in a second
achromat, where three stations of a MICROMEGAS type detector
operating in TPC mode form the kaon beam spectrometer (KABES)
(however, not used in the present analysis).

Downstream of the second achromat both beams follow the same path.
After passing the cleaning and the final collimators they enter the
decay volume, housed in a 114 m long cylindrical vacuum tank with a
diameter of 1.92 m for the first 65 m, and 2.4 m for the rest. With
$7\times 10^{11}$ protons per burst of $\sim 4.8$ s duration
incident on the target, the positive (negative) beam flux at the
entrance of the decay volume is $3.8\times 10^7$ ($2.6\times 10^7$)
particles per pulse, of which $5.7\%$ ($4.9\%$) are $K^+$ ($K^-$).
The $K^+/K^-$ flux ratio is about $1.8$. The fraction of beam kaons
decaying in the decay volume is about $22\%$.

The decay volume is followed by a magnetic spectrometer used for the
reconstruction of $K^\pm\to3\pi^\pm$ decays. The spectrometer is
housed in a tank filled with helium at nearly atmospheric pressure,
separated from the vacuum tank by a thin ($0.0031$ radiation
lengths) {\it Kevlar}-composite window. A thin-walled aluminium beam
tube of $\sim 16$ cm diameter traversing the centre of the
spectrometer (and all the other detectors) allows the undecayed beam
particles and the muon halo from decays of beam pions to continue
their path in vacuum. The spectrometer consists of four drift
chambers (DCH): two located upstream and two downstream of the
dipole magnet, which provides a horizontal transverse momentum kick
of about 120 MeV/$c$ to the charged particles. The DCHs have a shape
of a regular octagon with transverse size of about 2.8 m and
fiducial area of about 4.5 m$^2$. Each chamber is composed of eight
planes of sense wires arranged in four couples of staggered planes
(so called views) oriented horizontally, vertically, and along each
of the two orthogonal $45^\circ$ directions. The momentum resolution
of the magnetic spectrometer is $\sigma_p/p = 1.02\% \oplus
0.044\%p$ ($p$ is expressed in GeV/$c$), corresponding to a
reconstructed $3\pi$ invariant mass resolution of about
$1.7$~MeV/$c^2$.

The magnetic spectrometer is followed by a scintillator hodoscope
consisting of a plane of horizontal and a plane of vertical strips,
each plane is arranged in four quadrants (strip widths are 6.5 cm
for central counters and 9.9 cm for peripheral ones). The hodoscope
is in turn followed by a liquid krypton electromagnetic calorimeter
(LKr), a hadronic calorimeter (HAC), and a muon detector (MUV).

The $K^\pm\to3\pi^\pm$ decays are triggered with a two-level system.
At the first level (L1), the rate of $\sim 500$ kHz is reduced to
$\sim 100$ kHz by requiring coincidences of hits in the two planes
of the scintillator hodoscope in at least two quadrants. The second
level (L2) is based on a hardware system computing coordinates of
DCH hits from DCH drift times and a farm of asynchronous
microprocessors performing fast reconstruction of tracks and running
the decision taking algorithm. The L2 algorithm requires at least
two tracks to originate in the decay volume with the closest
distance of approach of less than 5 cm. L1 triggers not satisfying
this condition are examined further and accepted if there is a
reconstructed track that is not kinematically compatible with a
$\pi^\pm\pi^0$ decay of a $K^\pm$ having momentum of 60 GeV/$c$
directed along the $z$ axis. The resulting trigger rate is about 10
kHz.

The description of other components of the NA48 detector less
relevant for the present analysis can be found elsewhere\cite{fa99}.

NA48/2 collected data during two runs in 2003 and 2004, with
$\sim$50 days of efficient data taking each. About $18\times 10^9$
triggers were totally recorded.

\section{Measurement method}
The measurement method is based on comparing the reconstructed $u$
spectra of $K^+$ and $K^-$ decays $N^+(u)$ and $N^-(u)$. Given the
actual values of the slope parameters $g$, $h$, and $k$\cite{pdg}
and the precision of the current measurement, the ratio of $u$
spectra of $R(u)=N^+(u)/N^-(u)$ is in good approximation
proportional to $(1+\Delta g\cdot u)$, so $\Delta g$ can be
extracted from a linear fit to the ratio $R(u)$, and $A_g=\Delta
g/2g$ can be evaluated.

Charge symmetrization of the experimental conditions is to a large
extent achieved by using simultaneous and collinear $K^+$ and $K^-$
beams with similar momentum spectra. However, the presence of
magnetic fields in both the beam line (achromats, focusing
quadrupoles, etc.) and the magnetic spectrometer, combined with some
asymmetries in detector performance, introduces residual charge
asymmetries. In order to equalize the local effects on the
acceptance, the polarities of all the magnets in the beam line were
reversed during the data taking on an approximately weekly basis
(corresponding to the periodicity of SPS technical stops), while the
polarity of the spectrometer magnet was alternated on a more
frequent basis (approximately once per day in 2003 and once in 3
hours in 2004).

Data collected over a period with all the four possible setup
configurations (i.e. combinations of beam line and spectrometer
magnet polarities) spanning about two weeks of efficient data taking
represent a ``supersample'', which is treated as an independent and
self-consistent set of data for asymmetry measurement. Nine
supersamples numbered 0 to 8 were collected in two years of data
taking (supersamples 0--3 in 2003 and supersamples 4--8 in 2004).

Each supersample contains four sets of simultaneously collected
$K^+\to3\pi$ and $K^-\to3\pi$ samples corresponding to the four
different setup configurations (totally eight data samples). To
measure the charge asymmetry, exploiting the cancellations of
systematic biases emerging due to polarity reversals, the following
``quadruple ratio'' involving the eight corresponding $u$ spectra,
composed as a product of four $R(u)=N^+(u)/N^-(u)$ ratios with
opposite kaon sign, and deliberately chosen setup configurations in
numerator and denominator, is considered:
\begin{equation}
R_4(u) = R_{US}(u)\cdot R_{UJ}(u)\cdot R_{DS}(u)\cdot R_{DJ}(u).
\label{quad}
\end{equation}
In the notation for the $R(u)$ ratios, the indices $U$ ($D$) denote
to beam line polarities corresponding to $K^+$ passing along the
upper (lower) path in the achromats, respectively, while the indices
$S$ ($J$) represent spectrometer magnet polarities (opposite for
$K^+$ and $K^-$) corresponding to the ``even'' (i.e. the two
identical) pions being deflected to negative (positive) $x$, i.e.
towards the Sal\`eve (Jura) mountains, respectively. A fit of the
quadruple ratio~(\ref{quad}) with a function $f(u)=n\cdot(1+\Delta
g\cdot u)^4$ results in two parameters: the normalization $n$ and
the difference of slopes $\Delta g$. The normalization is sensitive
to the $K^+/K^-$ flux ratio, while $\Delta g$ is not.

The quadruple ratio technique logically completes the procedure of
magnet polarity reversal, and allows a three-fold cancellation of
systematic biases:
\begin{itemize}
\item due to spectrometer magnet polarity reversal, local detector
  biases cancel between $K^+$ and $K^-$ samples with decay products
  reaching the same parts of the detector in each of the four ratios
  $R(u)$ appearing in the quadruple ratio $R_4(u)$;
\item due to the simultaneous beams, global time-variable biases cancel
  between $K^+$ and $K^-$ samples in the product of $R_S(u)$ and $R_J(u)$ ratios;
\item due to beam line polarity reversal, local beam line biases,
  resulting in slight differences in beam shapes and momentum spectra,
  largely cancel between the $R_U(u)$ and $R_D(u)$ ratios.
\end{itemize}
Remaining systematic biases due to the presence of permanent
magnetic fields (Earth's field, vacuum tank magnetization) are
minimized by using azimuthally symmetric geometrical acceptance cuts
discussed in the next section.

The method is independent of the $K^+/K^-$ flux ratio and the
relative sizes of the samples collected with different magnet
configurations. However, the statistical precision is limited by the
smallest of the samples involved, so the balance of sample sizes was
controlled during the data taking. The result remains sensitive only
to time variations of asymmetries in the experimental conditions
which have a characteristic time smaller than corresponding field
alternation period, and in principle should be free of systematic
biases.

Due to the method described above, no Monte Carlo (MC) corrections
to the acceptance are required. Nevertheless, a detailed GEANT-based
MC simulation was developed as a tool for systematic studies,
including full detector geometry and material description,
simulation of time-variable local DCH inefficiencies, time
variations of the beam geometry and DCH alignment. A large MC
production was carried out, providing a sample of a size comparable
to that of the data ($\sim 10^{10}$ generated events).
\section{Data analysis}
Several stages of data compaction and filtering were necessary in
order to reduce the data volume from $\sim 200$ TB of raw data to
$1.23$ TB of the final sample, while reducing the number of events
in the sample to $4.87\times10^9$. The principal requirement of the
filtering algorithm is the presence of at least 3 reconstructed
tracks in the event, which is not passed by about $55\%$ of the
recorded triggers.

Tracks are reconstructed from hits in DCHs using the measured
magnetic field map of the spectrometer analyzing magnet rescaled
according to the recorded current. Three-track vertices compatible
with a $K\to3\pi$ decay topology are reconstructed by extrapolation
of track segments from the upstream part of the spectrometer back
into the decay volume, taking into account the stray magnetic fields
due to the Earth and magnetization of the vacuum tank, and multiple
scattering at the Kevlar window. The stray field correction is based
on a three-dimensional field map measured in the entire vacuum tank.
It reduces the observed azimuthal variation of the reconstructed
$3\pi$ invariant mass by more than an order of magnitude to a level
below 50 keV/$c^2$. The kinematics of the events is calculated using
the reconstructed track momenta and directions at the vertex.

Event selection includes rejection of trigger buffer
overflow\footnote{Trigger buffer overflow indicates that a
  corresponding event has an enhanced number of hits in DCHs and
  therefore saturates the L2 trigger processor leading to significant
  reduction of the efficiency.}
events, requirements on vertex charge, quality, and position (within
the decay volume, and laterally within the beam), limits on the
reconstructed $3\pi$ momentum: $54~{\rm GeV}/c<P_K<66~{\rm GeV}/c$
and invariant mass: $|M_{3\pi}-M_K|<9$~MeV/$c^2$ (the latter
condition corresponds to roughly five times the resolution). The
selection leaves a sample of $3.82\times 10^9$ $K_{3\pi}$ events
which is practically background free, as $K_{3\pi}$ is the dominant
three-track decay mode of the charged kaon.\vspace{2mm}

\noindent {\it Fine alignment of the magnetic spectrometer}
\vspace{2mm}

\noindent The transverse positions of DCHs and individual wires were
realigned every 2--4 weeks of data taking using data collected in
special runs in which muon tracks were recorded with no magnetic
field in the spectrometer. This allows an alignment precision of
$\sim30~\mu$m to be reached. However, time variations of DCH
alignment on a shorter time scale can potentially bias the
asymmetry, since an uncorrected shift of a DCH along the $x$ axis
leads to charge-antisymmetric mismeasurement of the momenta.  An
unambiguous measure of the residual misalignment is the difference
between the average reconstructed $3\pi$ invariant masses for $K^+$
and $K^-$ decays ($\Delta\overline M$). A 1 $\mu$m shift of a DCH
along the $x$ axis induces a mass shift of $\Delta \overline
M\sim1$~keV/$c^2$ (the proportionality factor varies among the
DCHs). Monitoring of $\Delta \overline M$ revealed significant (up
to 200 $\mu$m) movements of the DCHs between individual alignment
runs. Introduction of time-dependent corrections to the measured
momenta based on the observed $\Delta\overline M$ reduces the effect
on the slope difference by more than an order of magnitude to a
negligible level of $\delta(\Delta g)<0.1\times 10^{-4}$.

\vspace{2mm}

\noindent {\it Correction for beam geometry instabilities}
\vspace{2mm}

\noindent The most important feature determining the geometric
acceptance is the beam tube traversing the centres of all DCHs.
Moreover, the beam optics can control the average transverse beam
positions to $\pm1$~mm, leaving a sizable random charge-asymmetric
bias to the acceptance. In order to compensate for this effect,
inner radius cuts $R>11.5$~cm are introduced for the distances $R$
of pion impact points at the first and the last DCHs from the actual
average measured beam position in DCH1 and its extrapolation from
upstream of the magnet in DCH4. These cuts cost $12\%$ of the
statistics, leading to a sample of $3.36\times 10^9$ events.

The minimum distance of 11.5 cm is chosen to ensure that the region
of the beam tube and the adjacent central insensitive areas of the
DCHs are securely excluded by the cut. The average beam positions
are continuously monitored separately for $K^+$ and $K^-$ by
calculating the momentum-weighted centre of gravity of the three
pions in the planes of DCHs upstream of the magnet. A bias
introduced by the fact that the average beam positions are
themselves affected by the acceptance is negligible. In addition to
the time variation of the average positions, also the dependencies
of the beam position on kaon momentum ($\sim 1$ mm in the horizontal
plane, $\sim 1$ cm in the vertical plane) and time in spill ($\sim
1$ mm) are monitored and taken into account. A precision of
$100~\mu$m in the determination of beam position is sufficient to
reduce systematic effects to a negligible level.

Residual charge-asymmetric effects originate from permanent
(irreversible) magnetic fields in the decay volume coupling to
time-dependent DCH inefficiencies and beam migrations. The
corresponding fake asymmetries have been estimated not to exceed
$\delta(\Delta g)=0.2\times10^{-4}$.

\vspace{2mm}

\noindent {\it Trigger efficiency correction} \vspace{2mm}

\noindent Only charge-asymmetric trigger inefficiencies dependent on
$u$ can possibly bias the measurement. Inefficiencies of the trigger
components are measured as functions of $u$ using control data
samples from low bias triggers collected along with the main
triggers. Thus time variations of inefficiencies can be accounted
for and the statistical errors on the inefficiency measurements are
propagated into the final result. Control trigger condition for L1
efficiency measurement requires at least one coincidence of hits in
the two planes of the scintillator hodoscope. Control triggers for
L2 efficiency measurement are L1 triggers recorded regardless of the
L2 response. The statistics of each of the two control samples is
roughly 1\% of the main sample.

The inefficiency of the L1 trigger, due to hodoscope inefficiency,
was measured to be $0.9\times 10^{-3}$ and found to be stable in
time. Due to temporary malfunctioning of particular hodoscope
channels, some fractions of data are affected by higher inefficiency
(up to $7\times 10^{-3}$), the source of the inefficiency being
localized in space. This kind of inefficiency was reduced to the
common level (and symmetrized) in the selected data sample by
applying appropriate geometric cuts to the pion impact points on the
hodoscope surface for the relevant supersamples. This procedure cost
7.1\% of the statistics, and led to the final sample of $3.11\times
10^9$ events. Due to the time stability of the L1 inefficiency, no
correction is applied, and an uncertainty of $\delta(\Delta
g)=0.3\times 10^{-4}$, limited by the statistics of the control
sample, is attributed.

For the L2 trigger, corrections to $u$ spectra are introduced for
the rate-independent part of the inefficiency, which is
time-dependent due to instabilities of the local DCH inefficiencies
affecting the trigger more than the reconstruction due to lower
redundancy. The integral inefficiency for the selected sample is
normally close to $0.6\times 10^{-3}$, but some periods are affected
by higher inefficiency of up to 1.5\% (its sources not being
trivially localized in space). The L2 inefficiency correction to the
whole statistics amounts to $\delta(\Delta g)=(-0.1\pm0.3)\times
10^{-4}$, where the error is statistical due to the limited size of
the control sample. The symmetry of the rate-dependent part of the
inefficiency of $\sim 0.2\%$ was checked separately with MC
simulation of pile-up effects and a study of the dependence of the
result on the number of allowed accidental tracks. \vspace{2mm}

\noindent {\it Asymmetry fits and cross-checks} \vspace{2mm}

\begin{figure}[t]
  \includegraphics{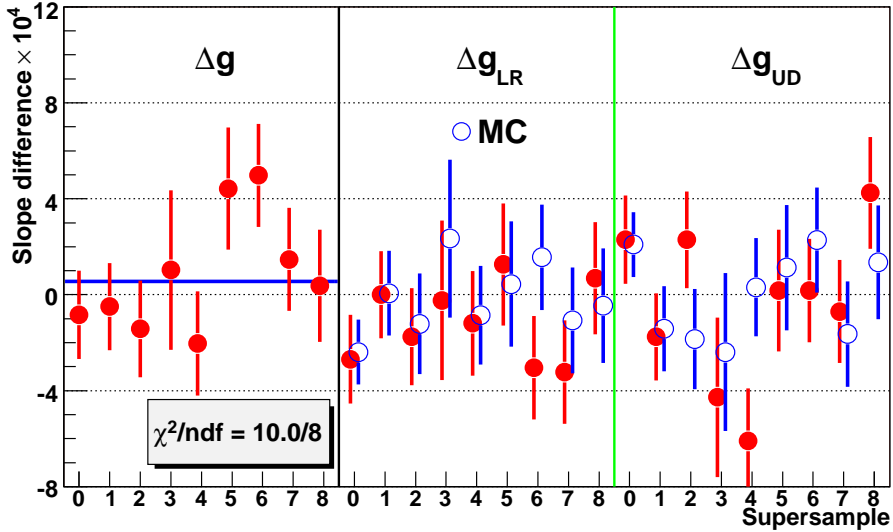}
    \vspace{70mm}
  \caption{\it (a) $\Delta g$ measurement in the four
supersamples; control
  quantities (b) $\Delta g_{LR}$ and (c) $\Delta g_{UD}$ corresponding
  to detector and beam line asymmetries which cancel in quadruple
  ratio, and their comparison to Monte Carlo.}
    \label{fig:stabplot}
\end{figure}

\noindent After applying the corrections described above, $\Delta g$
is extracted by fitting the quadruple ratio of the $u$ spectra
(\ref{quad}) independently for each supersample. The numbers of
events selected in each supersample, the ``raw'' values of $\Delta
g$ obtained without applying the trigger corrections and the final
values of $\Delta g$ with the L2 trigger corrections applied are
presented in Table~\ref{tab:stats}. The independent results obtained
in the nine supersamples are shown in Figure \ref{fig:stabplot}(a):
the individual measurements are compatible with a $\chi^2/{\rm ndf}
= 10.0/8$.

\begin{table}[tb]
\begin{center}
\begin{tabular}{r|r|r|r|r}
\hline Supersample & $K^+\!\to\pi^+\pi^+\pi^-\!\!$ &
$K^-\!\to\pi^-\pi^-\pi^+\!\!$
&$\Delta g\times 10^4$ & $\Delta g\times 10^4$\\
&decays in $10^6\!\!$&decays in $10^6\!\!$&raw~~~~~&corrected\\
\hline
0 & 448.0 & 249.7 & $ 0.5\pm1.4$ & $-0.8\pm1.8$\\
1 & 270.8 & 150.7 & $-0.4\pm1.8$ & $-0.5\pm1.8$\\
2 & 265.5 & 147.8 & $-1.5\pm2.0$ & $-1.4\pm2.0$\\
3 &  86.1 &  48.0 & $ 0.4\pm3.2$ & $ 1.0\pm3.3$\\
4 & 232.5 & 129.6 & $-2.8\pm1.9$ & $-2.0\pm2.2$\\
5 & 142.4 &  79.4 & $ 4.7\pm2.5$ & $ 4.4\pm2.6$\\
6 & 193.8 & 108.0 & $ 5.1\pm2.1$ & $ 5.0\pm2.2$\\
7 & 195.9 & 109.1 & $ 1.7\pm2.1$ & $ 1.5\pm2.1$\\
8 & 163.9 &  91.4 & $ 1.3\pm2.3$ & $ 0.4\pm2.3$\\
\hline
Total & 1998.9 & 1113.7 & $0.7\pm0.7$ & $0.6\pm0.7$\\
\hline
\end{tabular}
\end{center}
\vspace{-7mm} \caption{Statistics selected in each supersample and
the measured
  $\Delta g$: ``raw'' and corrected for L2 trigger inefficiency.} \label{tab:stats}
\end{table}

As a systematic check, to measure the size of the systematic biases
cancelling due to the quadruple ratio technique, two other quadruple
ratios of the eight $u$ spectra were formed. These are the products
of four ratios of $u$ spectra of same sign kaons recorded with
different setup configurations, therefore any physical asymmetry
cancels in these ratios, while the setup asymmetries do not. The
fake slope difference $\Delta g_{LR}$ introduced by global
time-variable biases does not cancel in ratios with opposite
spectrometer polarities and identical beam line polarities in
numerator and denominator, or equivalently, in the adopted notation
\begin{equation}
R_{LR}(u) = (R_{US}(u)\cdot R_{DS}(u)) / (R_{UJ}(u)\cdot R_{DJ}(u)).
\label{lr}
\end{equation}
Similarly, the fake slope difference $\Delta g_{UD}$ introduced by
the differences of the two beam paths does not cancel in ratios with
opposite beam line polarities and identical spectrometer polarities,
which look in the following way in the adopted notation:
\begin{equation}
R_{UD}(u) = (R_{US}(u)\cdot R_{UJ}(u)) / (R_{DS}(u)\cdot R_{DJ}(u)).
\label{ud}
\end{equation}
The measured fake slope differences $\Delta g_{LR}$ and $\Delta
g_{UD}$ in the nine supersamples are presented in Figure
\ref{fig:stabplot}(b) and (c) for both data and MC. The size of
these control quantities demonstrates that the cancellation of the
first-order systematic biases in (\ref{quad}), due to residual
time-variable imperfections in the apparatus, is at the level of a
few $10^{-4}$; therefore second order effects are negligible.
Moreover, the comparison with MC simulations shows that the sizes of
the apparatus asymmetries are well understood in terms of local
inefficiencies and beam optics variations.

\vspace{2mm}

\noindent {\it Limits for residual systematic effects} \vspace{2mm}

\noindent The measurement of pion momenta is based on the knowledge
of the magnetic field in the spectrometer magnet. The variation of
the current in the magnet can be monitored with a relative precision
of $5\times 10^{-4}$. Smaller variations are continuously controlled
with a precision of $\sim 10^{-5}$ by the deviation of the measured
charge-averaged kaon mass from the nominal value. A time-dependent
correction can be introduced by scaling the reconstructed pion
momenta symmetrically for positive and negative tracks. However,
this rather large momentum scale effect is charge symmetric by
design, due to the simultaneous beams. A conservative upper limit of
$\delta(\Delta g) = 0.1\times 10^{-4}$ for the corresponding
systematic uncertainty was obtained by comparing the results
evaluated with and without the correction.

In a considerable fraction of the selected events ($\sim 5\%$) at
least one of the pions undergoes a $\pi\to\mu\nu$ decay in the decay
volume and the spectrometer reconstructs the resulting muon. The
tails of the reconstructed $3\pi$ invariant mass distribution are
dominated by these events. Rejection of events with $\pi\to\mu\nu$
decays in MC analysis did not lead to any significant change of the
result. By varying the accepted $3\pi$ invariant mass interval in a
wide range (5--25 MeV/$c^2$) a conservative systematic uncertainty
of $\delta(\Delta g)=0.4\times 10^{-4}$ was attributed to effects
due to pion decays.

Taking into account that the composition of the beams is not charge
symmetric, event distortions caused by pile-up with products of
another kaon decay or a beam halo particle traversing the sensitive
region of the spectrometer is a potential source of systematic bias.
To study the pile-up effects, an accidental activity generator,
tuned using the measured beam and halo fluxes, composition and
kinematic distributions, was introduced into the MC, and a
production of $\sim10^7$ correlated pairs of an initial event and a
perturbed piled-up event was carried out. No charge-asymmetric
effects on $u$ distributions either in the reconstruction or in the
second level trigger were observed down to a level of $\delta(\Delta
g)=0.2\times 10^{-4}$, limited by MC statistics.

Biases due to resolution effects were studied by using various
methods of $u$ variable calculation from the measured track momenta
(including a 4C kinematic fit) differing in $u$ resolution as a
function of $u$ itself. The result is stable within $\delta(\Delta
g)=0.3\times 10^{-4}$.

Charge-asymmetric material effects have been found to be negligible
by evaluating the effect of hadronic interactions in the material in
front of and in the chambers taking into account the pion spectra.

A summary of the systematic uncertainties and trigger corrections is
presented in Table~\ref{tab:syst}.

\begin{table}[tb]
\begin{center}
\begin{tabular}{l|c}
\hline
Systematic effect & Correction, uncertainty $\delta(\Delta g)\times 10^4$\\
\hline
Spectrometer alignment        & $\pm0.1$\\
Acceptance and beam geometry  & $\pm0.2$\\
Momentum scale                & $\pm0.1$\\
Pion decay                    & $\pm0.4$\\
Pile-up                       & $\pm0.2$\\
Resolution and fitting        & $\pm0.3$\\
\hline
Total systematic uncertainty  & $\pm0.6$\\
\hline
Level 1 trigger               & $\pm0.3$\\
Level 2 trigger               & $0.1\pm0.3$\\
\hline
\end{tabular}
\end{center}
\vspace{-6mm} \caption{Systematic uncertainties and
  correction for level 2 trigger inefficiency.}
\label{tab:syst}
\end{table}
\section*{Conclusions}
The difference in the linear slope parameter of the Dalitz plot for
$3\pi^\pm$ decays of $K^+$ and $K^-$, measured with the full NA48/2
data sample, is found to be
\begin{equation}
\Delta g = g^+-g^- = (0.6 \pm 0.7_{stat.} \pm 0.4_{trig.} \pm
0.6_{syst.})\times 10^{-4}.
\end{equation}
Converted to the direct CP violating charge asymmetry in
$K^\pm\to3\pi^\pm$ decays using the PDG value of the Dalitz plot
slope $g=-0.2154\pm0.0035$\cite{pdg}, this leads to
\begin{equation}
A_g = (-1.3 \pm 1.5_{stat.} \pm 0.9_{trig.} \pm 1.4_{syst.})\times
10^{-4} = (-1.3\pm2.3)\times 10^{-4}.
\end{equation}

Analysis based on the full data sample has made a notable
improvement in the final uncertainty with respect to the published
analysis based on about a half of the data sample\cite{k3pi}. The
result has $\sim 17$ times better precision than the best
measurement before NA48/2, and the precision is still limited mainly
by the available statistics (the uncertainty due to the trigger
inefficiency is of statistical nature).

The result is compatible with the Standard Model predictions,
however due to its high precision it can be used to constrain
extensions of the Standard Model predicting enhancements of the
charge asymmetry.
\end{document}